\documentclass[aip,reprint,english]{revtex4-1}
\usepackage[T1]{fontenc}
\usepackage[latin9]{inputenc}
\usepackage{amsmath}
\usepackage{amssymb}
\usepackage{graphicx}
\pdfoutput=1
\makeatletter
\usepackage{epstopdf}
\usepackage{hyperref}
\hypersetup{
  linkcolor  = blue,
  citecolor  =blue,
  urlcolor   = blue,
  colorlinks = true,
}

\makeatother

\makeatletter
\g@addto@macro{\UrlBreaks}{\UrlOrds}
\g@addto@macro{\UrlBreaks}{%
\do\/\do\d%
}
\makeatother

\usepackage{babel}
\begin{document}

\title{Magneto-optical polarization rotation in a ladder-type atomic system
for tunable offset locking}

\author{Micha\l{} Parniak}

\email{michal.parniak@fuw.edu.pl}

\selectlanguage{english}%

\affiliation{Institute of Experimental Physics, Faculty of Physics, University
of Warsaw, Pasteura 5, 02-093 Warsaw, Poland}

\author{Adam Leszczy\'{n}ski}

\affiliation{Institute of Experimental Physics, Faculty of Physics, University
of Warsaw, Pasteura 5, 02-093 Warsaw, Poland}

\author{Wojciech Wasilewski}

\affiliation{Institute of Experimental Physics, Faculty of Physics, University
of Warsaw, Pasteura 5, 02-093 Warsaw, Poland}
\begin{abstract}
We demonstrate an easily tunable locking scheme for stabilizing frequency-sum
of two lasers on a two-photon ladder transition based on polarization
rotation in warm rubidium vapors induced by magnetic field and circularly
polarized drive field. Unprecedented tunability of the
two-photon offset frequency is due to strong splitting and shifting
of magnetic states in external field. In our experimental setup we achieve two-photon detuning of up to 700 MHz.\end{abstract}
\maketitle
Two-photon transitions allowing excitation to higher excited states are one of the moist promising candidates for  engineering light-atoms
interactions. They facilitate control of Rydberg atoms \cite{Peyronel2012} or ground-state
coherence \cite{Drampyan2009}, a two-color magnetooptical trap\cite{Wu2009,Yang2012a}, and a variety of other nonlinear wave-mixing
processes \cite{Parniak2015,Akulshin2009a,Kolle2012,Parniak2015d}. 

In numerous scenarios detuning the lasers from two-photon resonance by
a stable and well-controlled frequency offset is required. To obtain a steep
locking signal, one may modulate the absorption \cite{Akulshin2011}, transmission \cite{Moon2004}
or electromagnetically-induced transparency signals \cite{Bell2007,Abel2009}, or
use modulation transfer spectroscopy \cite{MartinezdeEscobar2015}.
Alternatievely, one can use Doppler-free polarization spectroscopy to
obtain the locking signal without modulation \cite{Wieman1976,Noh2012,Liao1976,Hamid2003}.
Nevertheless, tunability of the two-photon detuning requires additional
modulation at the offset frequency \cite{Yang2014}. 

In this Letter we present a modulation-free, easily tunable
scheme for frequency-sum stabilization of two lasers on a ladder transition.
By changing the magnetic field we are able to tune the dispersively
shaped locking signal around the two-photon resonance. The underlying principle is the polarization rotation of probe light induced by circular
polarization of drive field and external magnetic field.
Varying the magnetic field within a small range of $\pm$60 G offers unprecedented
tunability of over 1 GHz. Magnetic field has previously been used
in the two-photon variation of commonly employed dichroic atomic vapor laser
lock setup, but exclusievely to obtain the Doppler-free steep signal itself
\cite{Becerra2009}. The setup did not offer tunability of the two-photon
offset frequency. Tuning capability of magnetic field has only been
used so far in absorptive\cite{Dabrowski2015a} or Faraday anomalous dispersion filters\cite{Zielinska2012} based on atomic
vapors.

\begin{figure}
\includegraphics[scale=1.4]{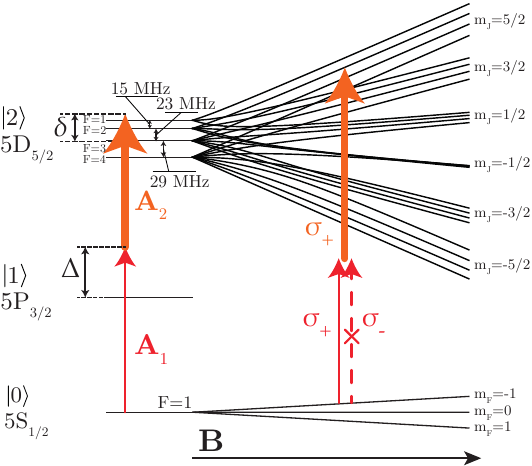}\caption{Atomic structure of $^{87}$Rb used in the experiment with a diagram of levels splitting
of the ground and highest excited states in magnetic field of up to
50 G. While the ground states remains well within the Zeeman regime, the $5\mathrm{D}_{5/2}$ manifold reaches the hyperfine Paschen-Back regime, since its hyperfine dipole constant $A_{5\mathrm{D}_{5/2}}=-7.44$ MHz\cite{Tai1975} is quite small. Circularly polarized drive field creates imbalance in susceptibilities for two circular polarizations of the probe field, giving rise to circular birefringence and in turn polarization rotation. \label{fig:levels}}
\end{figure}

Figure \ref{fig:levels} shows atomic levels involved in the locking
scheme. Linearly polarized probe field $\mathbf{A}_{1}$ at 780 nm
is far-detuned from the single-photon transition $5\mathrm{S}_{1/2}$,
$F=1$ $\rightarrow$ $5\mathrm{P}_{3/2}$ of $^{87}$Rb. Respective
detuning $\Delta$ is larger than both the hyperfine splitting of
$5\mathrm{P}_{3/2}$ state and the Doppler FWHM linewidth of approx. 635
MHz. Circularly polarized drive field $\mathbf{A}_{2}$ at
776 nm is detuned from the $5\mathrm{P}_{3/2}\ \rightarrow\ 5\mathrm{D}_{5/2}$
transition by $\delta-\Delta$, so the two lasers have combined nonzero two-photon detuning
$\delta$ from the line centroid. Counter-propagating configuration
results in nearly perfect cancellation of Doppler broadening for the
two-photon transition. Magnetic states are split in static magnetic
field $\mathbf{B}=B_{z}\mathbf{e}_{z}$ (along the propagation axis
$z$ of probe light).

Polarization rotation occurs due to birefringence induced by circularly polarized drive field and applied external magnetic field. Magnetic field also enables tuning, as it shifts atomic levels. In turn, the locking signal at different frequencies is obtained.

To describe the locking signal lineshapes we first adopt the model of an
isolated three level atom in ladder configuration. The atom is probed by
field of amplitude $\mathbf{A}_{1}$ and driven by field of
amplitude $\mathbf{A}_{2}$.  The susceptibility for the probe field corresponding to the steady-state solution of Maxwell-Bloch equations reads:

\begin{equation}
\chi=-\frac{{N}}{\hbar\epsilon_{0}}\frac{{|\mathbf{d}_{01}|^{2}}}{\Delta+i\Gamma/2-|\Omega_{2}|^{2}/4(\delta+i\gamma/2)},\label{eq:chi-exact}
\end{equation}

where $N$ is the atom number density and $\Omega_{2}=\mathbf{A}_{2}\cdot\mathbf{d}_{12}/\hbar$
is the Rabi frequency. Dipole moments of respective transitions are
given by $\mathbf{d}_{01}$ and $\mathbf{d}_{12}$ and relaxation rates of states $|1\rangle$
and $|2\rangle$ are given by $\Gamma=2\pi\times6.06$~MHz\cite{Steck} and $\gamma=2\pi\times0.66$
MHz \cite{Sheng2008}, respectively. For the single-photon detuning $\Delta\gg\Omega_{2},\Gamma$, we obtain an expression with two terms:

\begin{equation}
\chi=-\frac{{N}}{\hbar\epsilon_{0}}\left(\frac{{|\mathbf{d}_{01}|^{2}|\mathbf{d}_{12}|^{2}|\mathbf{A}_{2}|^{2}}}{4\hbar^{2}\Delta^{2}(\delta+i\gamma/2)}+\frac{{|\mathbf{d}_{01}|^{2}}}{\Delta}\right)\label{eq:chi-single}.
\end{equation}

The first term is resonant and dominates around $\delta=0$. The second,
linear dispersion term is very slowly-varying. Consequently, for our
considerations we may drop the latter and focus on the first, two-photon
term. 

To calculate the susceptibility for the atom having a rich hyperfine structure,
we sum over all possible sublevels $|n,\alpha\rangle$ (shifted in
frequency by $\omega_{n,\alpha}$) of each manifold $n$, thus taking into
account all possible two-photon transitions from $|0,\alpha\rangle$
to $|2,\alpha''\rangle$ : 

\begin{multline}
\chi_{q_{1},q_{2}}=-\frac{{N}}{4\hbar^{3}\epsilon_{0}}\\
\sum_{\alpha,\alpha',\alpha''}\frac{|\langle2,\alpha''|\hat{d}_{q_{2}}|1,\alpha'\rangle\langle1,\alpha'|\hat{d}_{q_{1}}|0,\alpha\rangle|^{2}|\mathbf{A}_{2}|^{2}}{(\Delta-\omega_{0,\alpha}+\omega_{1,\alpha'})^{2}(\delta-\omega_{0,\alpha}+\omega_{2,\alpha''}+i\gamma/2)},\label{eq:alpha-sum}
\end{multline}

where $\omega_{n,\alpha}$ is the shift of a particular sublevel $|n,\alpha\rangle$
from the centroid, $\hat{d}_{q}$ is the dipole moment operator and
light polarizations of fields $\mathbf{A}_{1}$ and $\mathbf{A}_{2}$
are given by $q_{1}$ and $q_{2}$. Since the single-photon detuning
$\Delta$ is far off-resonant we may neglect hyperfine and magnetic
splitting of the intermediate state and drop the $\omega_{1,\alpha'}$
term in the above expression. This simplifies the summation, since now
the summation over intermediate states indexed by $\alpha'$ may be
done separately. To find the dipole matrix elements, we first solve
the eigenproblem for the Hamiltonian in $|n,m_I,m_J\rangle$ basis for each
manifold $n$:

\begin{equation}
\hat{{H}}_{n}=\mu_{B}g_{n,J}B_{z}\hat{J}_{n,z}+A_{n}\mathbf{\hat{I}}_{n}\cdot\hat{\mathbf{J}}_{n},
\end{equation}

where $\mu_{B}$ is the Bohr magneton, $g_{n,J}$ is the Land\'e
factor, $A_n$ is the hyperfine dipole coupling constant and $\hat{\mathbf{I}}_{n}$ and $\hat{\mathbf{J}}_{n}$ are
nuclear and electronic angular momentum operators, respectively, for
hyperfine manifold $n$. We find decomposition of $|n,\alpha\rangle$
in $|n,m_I,m_J\rangle$ basis and respective energy shifts of levels $\hbar\omega_{n,\alpha}$.
Using this decomposition we are able to transform the dipole matrix
to Hamiltonian eigenbasis.

\begin{figure}
\centering{}\includegraphics[bb=0bp 0bp 180bp 137bp,scale=1.25]{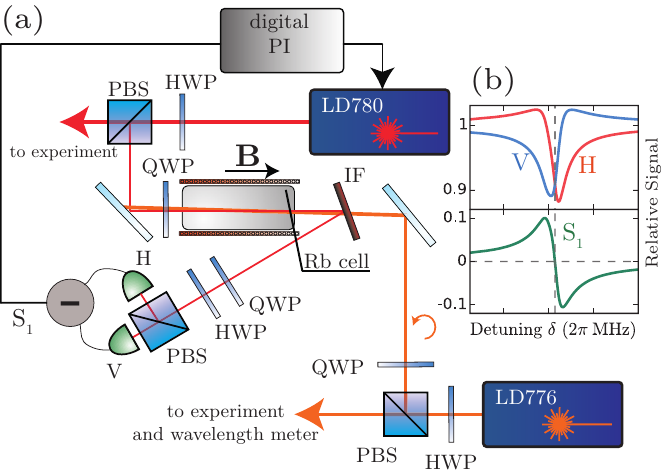}\caption{(a) Experimental configuration of the locking setup and (b) exemplary normalized
signals H and V registered with two photodiodes of a balanced detector
and the difference of the two signals $\mathrm{S}_{\mathrm{1}}$ demonstrating
dispersive shape for magnetic field of 24 G. \label{fig:experimental}}
\end{figure}

Finally, transmitted light intensities are calculated in circular polarization basis. Here, we give our result for the difference
signal, being proportional to the normalized Stokes parameter $\mathrm{S}_{1}=(\langle E_x^2\rangle-\langle E_y^2\rangle)/(\langle E_x^2\rangle+\langle E_y^2\rangle)$ of the probe field:
\begin{equation}
\mathrm{S}_{\mathrm{1}}\propto|\mathbf{A}_{1}|^{2}(\mathrm{Re}\chi_{1,q_{2}}-\mathrm{Re}\chi_{-1,q_{2}}).
\end{equation}

This expression, being the difference of real, dispersive parts of
susceptibilities for two circular polarizations is also proportional
to polarization rotation angle. See reference for Mathematica notebook
containing implementation of the above calculations\cite{Parniak2016}.

In our experimental implementation, we observe polarization rotation
of the probe light using a polarizing beamsplitter (PBS) and a balanced
detector, as shown in Fig.~\hyperref[fig:experimental]{\ref*{fig:experimental}(a)}. Exemplary shapes
of signals registered by the two photodiodes, exhibiting both dispersive
and absorptive behavior, are shown in Fig.~\hyperref[fig:experimental]{\ref*{fig:experimental}(b)}.
Their difference shows purely dispersive behavior, with a steep slope
in the center of resonance. To stabilize the laser frequency, we feed
the locking signal to a digital proportional-integral
(PI) controller, which subsequently adjusts current in the 780 nm
laser diode (LD780, Toptica DL100 DFB). The 776 nm laser (LD776, Toptica
DL100 ECDL) is stabilized using a commercial wavelength meter (Angstrom
HighFinesse WS7), which recently proved to be very suitable for laser frequency
stabilization \cite{Saleh2015}. In the 5-cm-long rubidium vapor cell
the two beams counter-propagate, having $1/e^{2}$ diameters of 500
$\mu$m. Maximum drive power is 50 mW, while the probe power is
500 $\mu$W. Probe light is separated using an interference filter
(IF, Thorlabs FBH780-10), tilted to reflect 780 nm light and transmit
776 nm light. Additional quarter-wave plates (QWP) before and after
the cell are used to compensate for the birefringence of cell windows and
the interference filter. Half-wave plate (HWP) is used to adjust balance
of the detector. Drive field $\sigma_{-}$ polarization is set using
a single QWP.

For the experimental proof-of-principle demonstration we choose a
transition from the ground state $5\mathrm{S}_{1/2},F=1$. For the
simplicity of notation from now on we measure the single-photon detuning
$\Delta$ from $5\mathrm{S}_{1/2},F=1$ state, just as in Fig. 1.
Note that magnetic splittings of both the ground state and the intermediate state are insignificant
when compared to the single-photon detuning $\Delta$, and consequently
in Eq. \ref{eq:alpha-sum} we take $(\Delta-\omega_{0,\alpha}+\omega_{1,\alpha'})^{2}\approx\Delta^{2}$
for each $\alpha$ and $\alpha'$.

\begin{figure}
\includegraphics[scale=0.72]{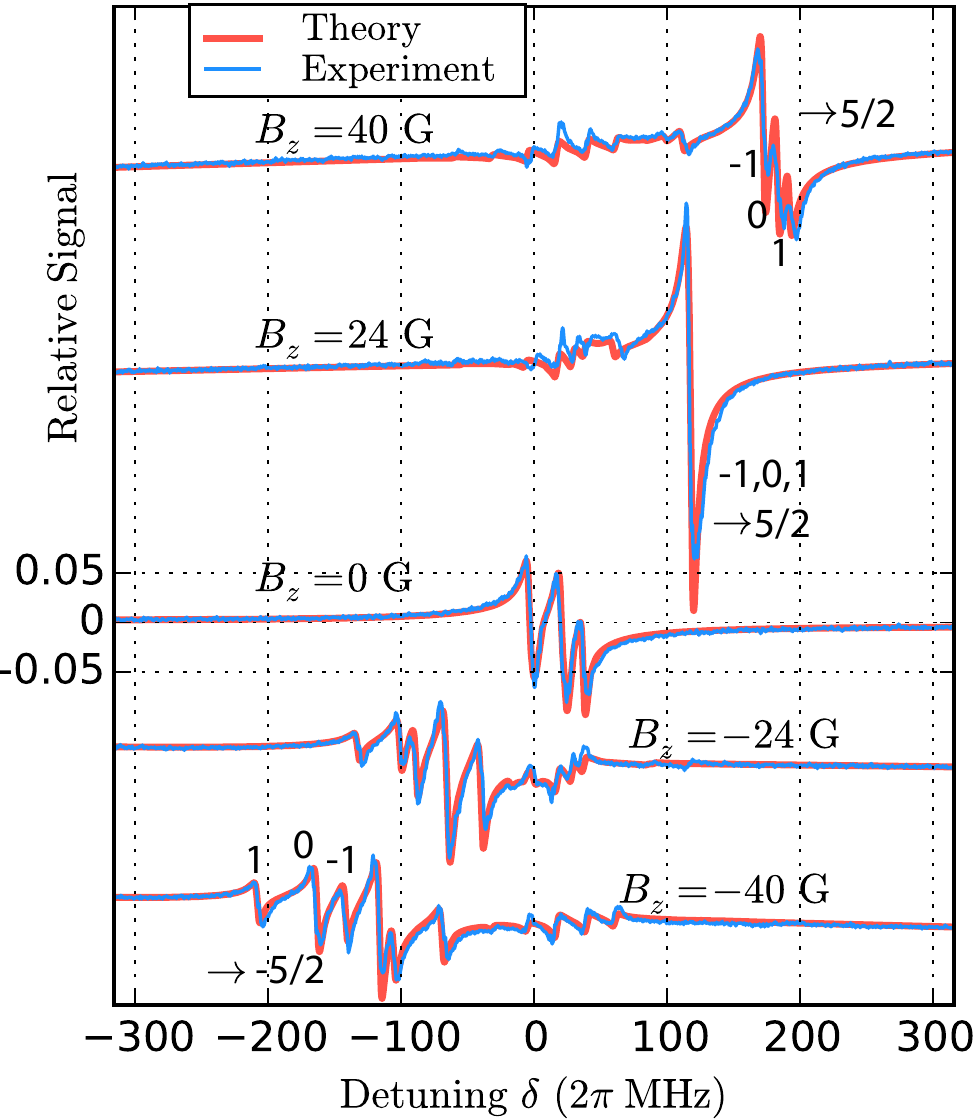}\caption{Exemplary locking signals normalized to sum of signals detected by
two photodiodes H and V for different magnetic fields $B_{z}$, $\Delta=2\pi\times4$ GHz and $I_{2}=240$ mW/mm$^{2}$. Two-photon resonances $F=1;\ m_F=-1,0,1 \rightarrow m_J=5/2$ are marked. \label{fig:signals}}
\label{fig:shapes}
\end{figure}
\begin{figure}
\includegraphics[scale=0.7]{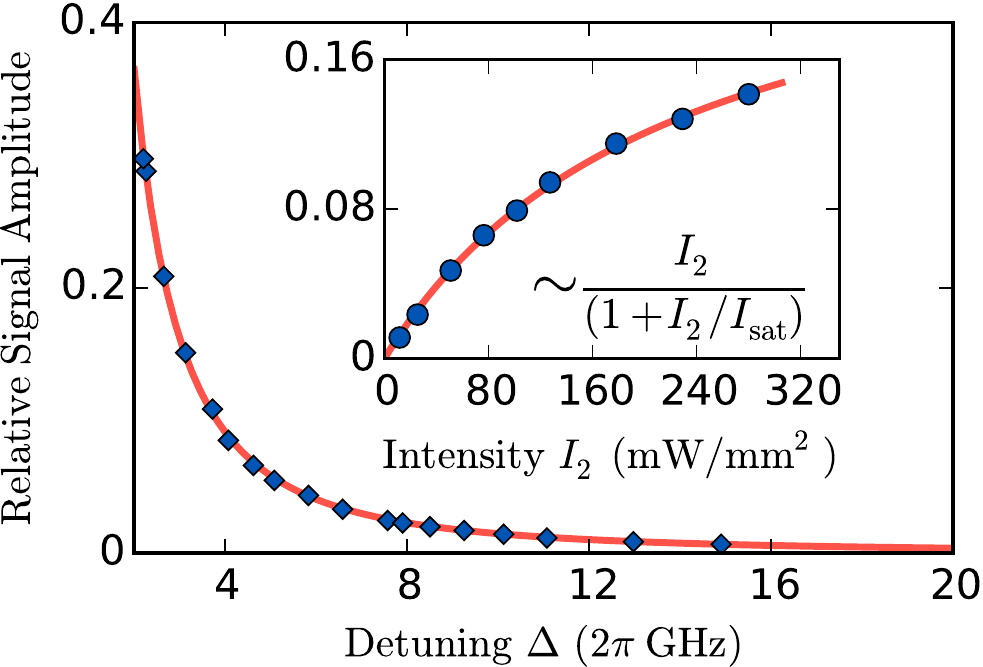}\caption{Locking signal peak-to-peak amplitude for $B_{z}=24$ G and drive
field intensity $I_{2}=240$ mW/mm$^{2}$ measured as a function of single-photon
detuning $\Delta$ (data points) demonstrating expected inverse-$\Delta^{2}$ dependance (solid line).
Inset shows dependance of the signal amplitude on drive field intensity
$I_{2}$ for $\Delta=2\pi\times3$ GHz.\label{fig:pkpampl} }
\label{fig:fits}
\end{figure}

In the first measurement we fix the detuning of 776 nm laser and scan
the detuning of 780 nm laser around the two-photon resonance. This
situation corresponds to virtually constant single-photon detuning $\Delta$ and varying two-photon detuning $\delta$. For different magnetic
fields we obtained shifted signals. For magnetic fields exceeding 15 G $m_{J}$ is a good quantum number describing magnetic states
of $5\mathrm{D}_{5/2}$ manifold. Consequently, lines that most rapidly
shift in magnetic field correspond to highest $m_{J}=5/2$, while
the barely shifted lines in the middle correspond to $m_{J}=1/2$
states. Splitting of the ground state only slightly influences shifting,
but has significant influence on signal shape. With our theoretical
model we are able to predict these shapes, and as seen in Fig.~\ref{fig:signals}
we obtain excellent conformity of theoretical prediciton and experimental data. The
relaxation rate of highest excited state $\gamma$ was adjusted for
additional broadening (mainly transit-time broadening and residual Doppler broadening\cite{Moon2014b,Suppl}) and found to
be $2\pi\times4.2$ MHz. Finally, note that instead of changing the sign of magnetic
field, one may equivalently change drive field polarization to the opposite.

Apart from the two-photon rotation, we also observed background signal
from single-photon Faraday effect at $|0\rangle\rightarrow|1\rangle$
transition. This effect can easily be compensated by rotating the
HWP.

\begin{figure}
\includegraphics[scale=0.72]{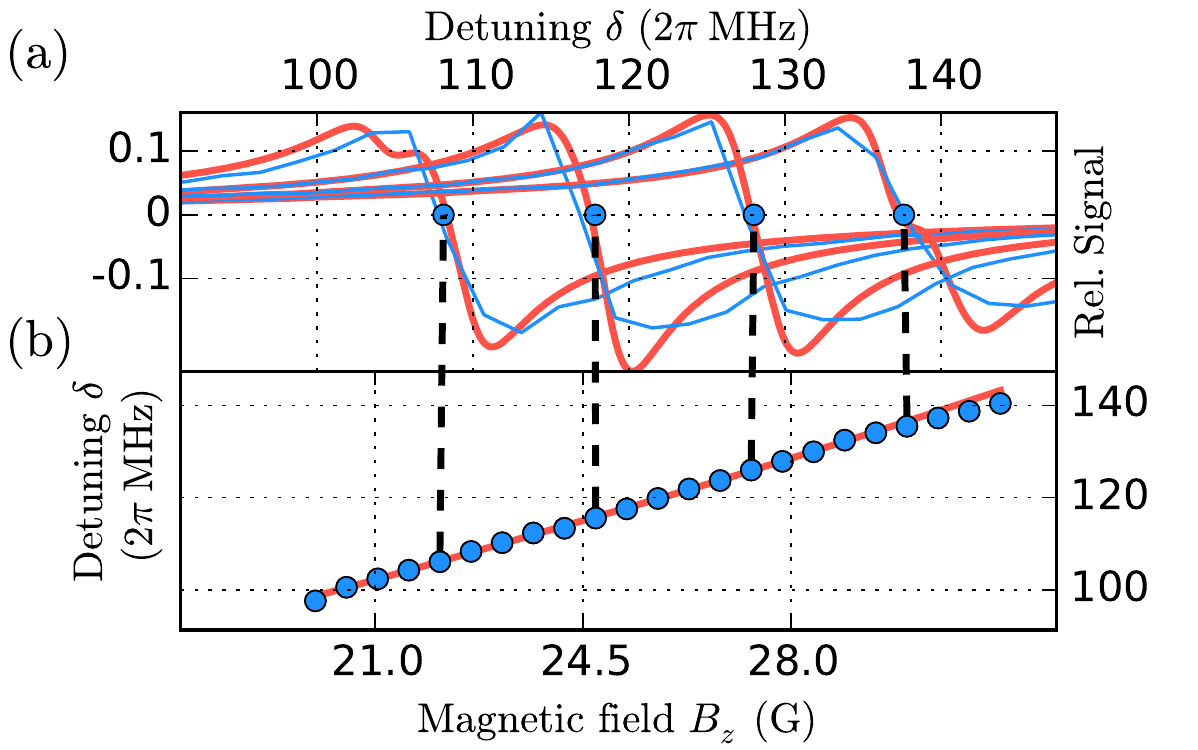}\caption{Tuning of the steep locking signal obtained around $B_{z}=24$ G: (a)
exemplary signals for $B_{z}=$22, 24.5, 27.3 and 29.7 G (from left to right), (b) zero-crossing
of the signal with predicted linear dependance on detuning. This line corresponds to $F=1;\ m_F=-1,0,1 \rightarrow m_J=5/2$ two-photon resonances. \label{fig:tuning}}
\label{fig:magtuning}
\end{figure}

At $B_{z}$ around 24 G we obtain enhanced signal, corresponding to $F=1 \rightarrow m_J=5/2$ transitions triplet.
Enhancement is due to constructive interference of contributions from different magnetic ground-state
levels assisted by compensation of splittings of initial and final states. In Fig. \ref{fig:pkpampl} we changed the single photon detuning
$\Delta$ and observed expected inverse-$\Delta^{2}$ dependance (see Eq. \ref{eq:chi-single}). In our setup,
the signal to electronic noise ratio is high enough for locking at single-photon detunings up to $\Delta=20$ GHz, where we obtain $\mathrm{SNR}=8$. We consistently found the signal to scale linearly with probe field intensity. The signal scales linearly with the
drive field intensity for higher single-photon detunings as well, but closer
to resonance the signal saturates \cite{Suppl} as shown in the inset of Fig. \ref{fig:pkpampl}.
\begin{figure}
\includegraphics[scale=0.72]{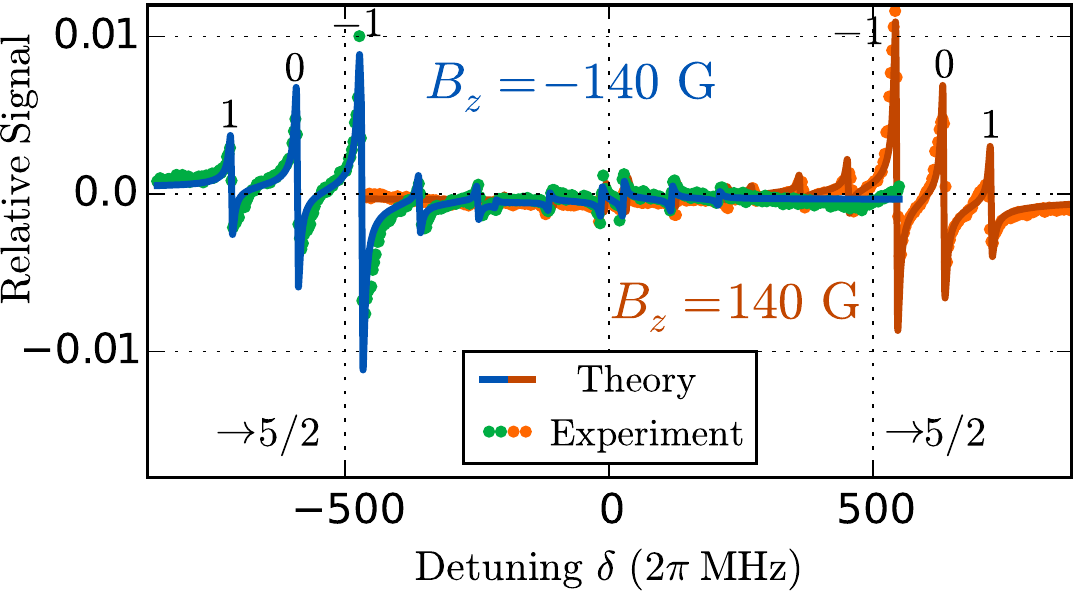}\caption{Calculated error signal for $I_{2}=240$ mW/mm$^{2}$, $\Delta=2\pi\times8$
GHz and two oposite large values of magnetic field demonstrating tunability
of two-photon offset frequency of over $1$ GHz. Strongest marked resonances correspond to $m_F=-1,0,1\rightarrow m_J=5/2$ transitions.\label{fig:largeb}}
\end{figure}

Finally, we vary the magnetic field around 24 G and observe linear
tuning of the zero-crossing of the signal. Four exemplary shapes are shown in Fig. \hyperref[fig:tuning]{\ref*{fig:tuning}(a)}. Linear dependance of the zero-crossing point on magnetic field is illustrated
in Fig. \hyperref[fig:tuning]{\ref*{fig:tuning}(b)}. We found the proportionality constant
to be $2\pi\times4.06$~MHz/G. This gives us an estimate on required
stability of magnetic field. For example, for 10 kHz stability we
require magnetic field instability to be lower than 3 mG, maximum
drive laser power instability of 1\% at 50 mW drive power and temperature instability less than 0.5$^\circ$C, which
are within the reach of current technical capabilities.

We envisage
that higher magnetic field allows very broad tunability \cite{Suppl}. Fig.~\ref{fig:largeb}
shows the signal calculated and measured for blue-detuning $\Delta=2\pi\times8$~GHz and $B_{z}=140$ G and -140 G, demonstrating possibility of locking at
two-photon detuning $\delta$ from -$2\pi\times700$ MHz to $2\pi\times700$
MHz. 

In conclusion, we have proposed and realized a setup enabling modulation-free
tunable offset locking at a two-photon ladder transition. Tunability
is achieved thanks to the splitting of the highest excited state in constant external
magnetic field. Even though polarization spectroscopy is widely used,
it is the first proposal of using magnetic field to tune the two-photon
offset frequency. At low magnetic fields we obtain signal insusceptible
to environmental perturbations. The wide range of tuning does not require any modulation.
We note that configuration of fields may be easily changed in our
setup, as well as in theoretical model, to use the $\mathbf{A}_{1}$
field on $|0\rangle\rightarrow|1\rangle$ transition as drive field.
This would provide the feedback for laser coupled to $|1\rangle\rightarrow|2\rangle$
transition. Another method would be to simply supply the error signal as feedback to the 776 nm laser\cite{Abel2009,Becerra2009}.
The setup we demonstrated could be also used to stabilize the two-photon
detuning when the excitation is done in a frequency-degenerate scheme
(e.g. by 778 nm light in case of $5\mathrm{S}_{1/2}\rightarrow5\mathrm{D}_{5/2}$
transition) \cite{Brekke2015}, possibly providing even narrower line
due to perfect cancellation of Doppler-broadening, or alternatievely at transitions
including telecommunication wavelengths \cite{MartinezdeEscobar2015}.

This work has been supported by Polish Ministry of Science and Higher
Education ``Diamentowy Grant'' Project No. DI2013 011943, National
Science Center Grant No. DEC-2011/03/D/ST2/01941 and by the Seventh Framework Programme PhoQuS@UW Project (Grant Agreement No. 316244).  We acknowledge generous support of T. Stacewicz and K. Banaszek, as well as careful proofreading of the manuscript by M. Jachura and M. D\k{a}browski.
\bibliographystyle{apsrev4-1}
\bibliography{bibliografia}

\end{document}